\documentclass[aps,prb,twocolumn,showpacs]{revtex4}
\usepackage{graphicx}% Include figure files 
\begin{document} 
%\twocolumn[\hsize\textwidth\columnwidth\hsize\csname @twocolumnfalse\endcsname 

\title{Disorder induced local density of states oscillations on narrow 
Ag(111) terraces} 

\author{Karina Morgenstern and Karl-Heinz Rieder} 
\address{Institut f\"ur Experimentalphysik, FB Physik, Freie Universit\"at Berlin, 
Arnimallee 14, D-14195 Berlin, Germany} 
\author{Gregory A. Fiete} 
\address{Kavli Institute for Theoretical Physics, University of California, Santa Barbara, CA 93106, USA} 
\date{\today} 

\begin{abstract} 

The local density of states of Ag(111) has been probed in detail on
disordered terraces of varying width by dI/dV-mapping with a scanning
tunneling microscope at low temperatures.  Apparent shifts of the
bottom of the surface-state band edge from terrace induced confinement
are observed. Disordered terraces show interesting contrast reversals
in the dI/dV maps as a function of tip-sample voltage polarity with
details that depend on the average width of the terrace and the
particular edge profile.  In contrast to perfect terraces with
straight edges, standing wave patterns are observed parallel to the
step edges, i.e. in the non-confined direction. Scattering
calculations based on the Ag(111) surface states reproduce these
spatial oscillations and all the qualitative features of the standing
wave patterns, including the polarity-dependent contrast reversals.

\end{abstract} 
 %Scanning tunneling microscopy (including chemistry induced with STM) 
%] 
%\newpage 
\maketitle

\section{Introduction} 

The fcc(111) surfaces of noble metals exhibit a large sp-band gap in 
the projected bulk band structure along the $\Gamma$-L line. 
\cite{shockley39,forstmann70} These gaps reach down below the Fermi 
energy at the center of the surface Brillouin zone and support an 
occupied free electron surface-state.\cite{kevan89,roos89} Electrons 
occupying these Shockley-type surface-states form a two-dimensional 
nearly free electron gas parallel to the surface. 
\cite{gartland75,heimann77,zangwill88} Electrons are confined to the 
vicinity of the top layer by the vacuum barrier on one side and a band 
gap in the bulk states on the other side.

Scanning tunneling microscopy (STM) and spectroscopy (STS) are 
particularly sensitive tools to study these surface-states that 
dominate in the zone between the tip and the surface. \cite{tersoff85} 
These techniques map the local electronic density of states (LDOS) in 
real space. STM has provided real space observation of electrons in 
surface-states and of their interactions with adsorbates, 
\cite{crommie93_nature,crommie93_science} steps, 
\cite{crommie93_nature,crommie93_science,avouris94,sanchez95,garcia95_APA,ortega00}, 
and other structures 
\cite{crommie93_nature,crommie93_science,li97,li98} in the form of 
standing wave patterns. These spatial oscillations are quantum 
mechanical interference patterns caused by the scattering of electrons 
in the two-dimensional electron gas from defects. The 
standing wave patterns contain information 
about the surface-state dispersion and they give insight into the 
interaction between surface-state electrons and scattering sites on 
the surface. \cite{crommie93_nature,crommie93_science}

STS images of the local differential tip-surface conductance (called
dI/dV maps) have been used to investigate how the surface-state
interacts with a single straight atomic step
\cite{jeandupeux99,buergi99} and with a pair of adjacent straight
steps.\cite{avouris94,buergi99} From the measurement of spatial
oscillation periods normal to the step it was possible to extract
Fermi wave vectors \cite{sprunger97} and complete energy dispersion
curves.\cite{buergi00} Thus, the surface-state on clean, disorder-free
systems is well characterized. However, growth and catalysis, which
are both influenced by the electronic structure of the surface,
\cite{repp00,knorr02,morgenstern04} also occur on surfaces where
disorder is present. In particular, the LDOS modulations on irregular step
arrays\cite{morgenstern01} can have a dramatic impact on these
processes.  Therefore, it is important to understand and to be able to
predict the electronic structure on disordered terraces of varying
width limited by steps with irregular edges.  A study of the
electronic structure of such terraces is the focus of this paper.

In a recent publication, we have investigated the dependence of the 
LDOS near the Fermi-energy, E$_F$, on the width of 
terraces on Ag(111) at 7 K.\cite{morgenstern02} We showed that the {\em apparent} surface-state 
band minimum 
shifts monotonically towards the bulk $E_F$  with 
decreasing terrace width leading to a 
depopulation at a mean terrace width $L=3.2$ nm, 
in quantitative agreement with a Fabry-P\'{e}rot model with energy 
dependent finite asymmetric reflectivity of the step edges. In 
addition, a switch from confinement by terrace modulation to step 
modulation was observed at $\lambda_F/2$ ($L$=3.7 nm), half 
of the Fermi wavelength on an infinite terrace. 
This switch occurs on Cu(111) surfaces also at $\lambda_F/2$ as 
proven by photoemission. \cite{ortega00}

In this paper, we present STS data revealing spatial oscillations of
the local density of states on Ag(111) for narrow disordered terraces
of varying width up to 8.5 nm. Apparent band edge shifts of the
surface-state onset (as reported in Ref.~[\onlinecite{morgenstern02}])
lead to conductance variations on terraces of different width at the
same energy. A standing wave pattern also evolves parallel to the
steps, i.e.\ in the non-confined direction. Its wavelength follows the
dispersion relation of an electron in the clean, infinite
surface-state. In order to understand the wave pattern, theoretical
simulations based on scattering theory are performed for disordered
terraces. These calculations successfully reproduce the observed
oscillations parallel to the terraces as well as interesting
tip-sample voltage polarity dependent effects.  

This paper is organized in the following way. In Sec.~\ref{sec:exp} we
discuss the details of the experiment, including sample preparation,
the types of measurements made, and the main features observed in the
data. In Sec.~\ref{sec:theory} we discuss the theoretical model used
to interpret the experiments and its assumptions. We also present the
results of model calculations and discuss their connection with
experiment. Finally, in Sec.~\ref{sec:conclusions} we give the main
conclusions of our work.

\section{Experiment} 
\label{sec:exp} 

\subsection{Sample preparation and data acquisition} 

The experiments have been performed in ultrahigh
vacuum with a low temperature scanning tunneling microscope that
operates at temperatures between 6 and 300 K. The single crystalline
Ag(111) surface is cleaned by sputtering and annealing cycles. The
surface separates into large flat terraces and step bunches with an
average terrace width of 4 nm.\cite{morgenstern01} In this paper, we
investigate the spatial oscillations of the LDOS in these stepped
regions. Measurements are performed at 7 K.

The topographic images are taken in constant current mode. dI/dV 
spectra are recorded with the lock-in technique; the ac tunnel current is 
driven by a 4 mV signal added to the junction bias with a frequency of 
327.9 Hz, 381.8 Hz, or 738.1 Hz. Voltages are applied to the sample 
with respect to the tip. Thus, for a negative voltage the occupied 
side of the spectrum is probed. dI/dV maps are recorded 
simultaneously with the topographic images at the same voltage by 
recording the lock-in signal at each pixel. It is not possible to run 
the STM in constant height mode in imaging a step array of more than 1 
nm change in height. Thus, the constant-current image recorded 
simultaneously with the dI/dV map is used for an adjustment in 
tip-sample separation. This leads to a superposition of the dI/dV 
signal with a variance in height at the step edges.

A dI/dV map of 256x256 pixels is recorded in about 80 minutes. Series 
of dI/dV maps therefore requires a high stability STM, which our 
system provides. Typically, the drift is less than 0.3 nm during a 
series of 10 dI/dV maps, i.e.\ 13 hours. Major drift problems arise from 
the increasing temperature at the end of helium evaporation. 
Therefore, measurement series on the same spot of the crystal are 
limited by the helium amount available in the bath cryostat.

\subsection{Experimental results}

\begin{figure} %fig1 
\includegraphics{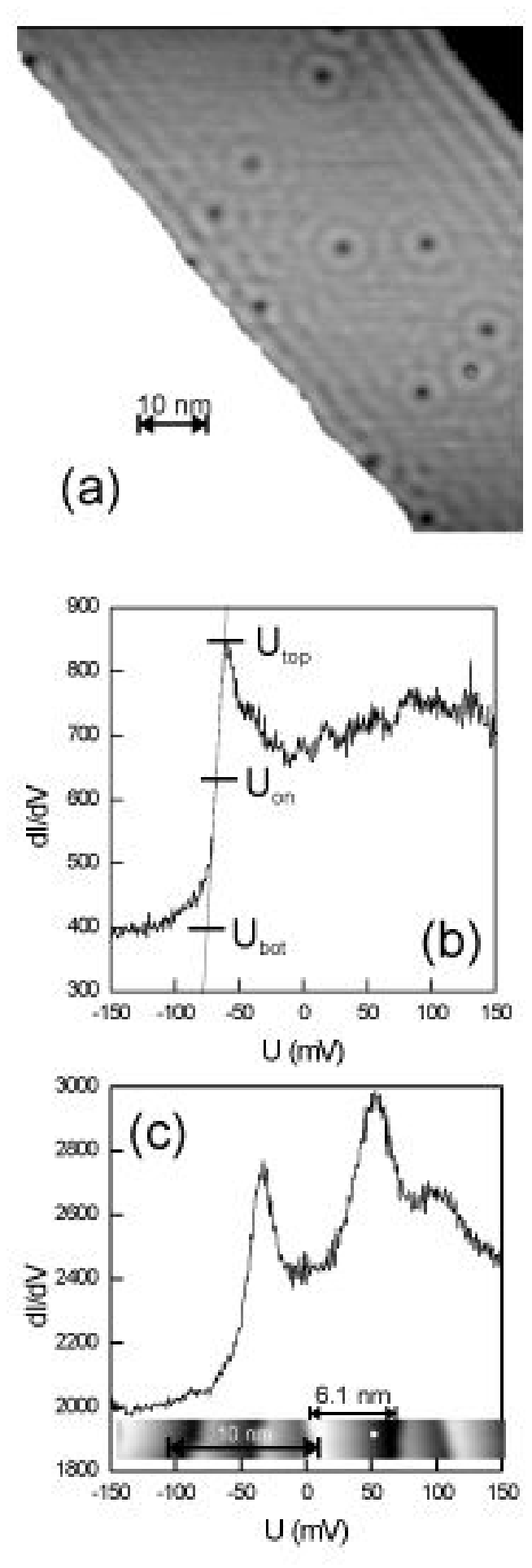} 
\caption[]{(a) Real space STM image taken at 7 K showing spatial oscillations of electrons in the Ag(111) surface-state 
on a large terrace 51 nm in width. Circularly symmetric oscillations in the LDOS are seen around point impurities. The tip-sample voltage bias is $U_{bias}$=32 mV with a tunneling current $I_{tun}$=0.43 nA. 
(b) The surface state onset is indicated in the dI/dV spectrum on the large terrace shown in (a). Here $U_{on}=(U_{top}+U_{bot})/2$ is taken as the value of the onset. 
(c) The dI/dV spectrum on the edge of a narrow terrace of 6.1 nm width (indicated by a white dot on the inset STM image) shows two peaks in intensity. These peaks can be related to the onset of the lowest transverse mode of the terrace (first peak $\approx -50$ mV) and the first excited tranverse mode (second peak $\approx 50$ mV). Here $U_{bias}$=-137 mV and $I_{tun}$=0.34 nA. 
\label{large_terrace}} 
\end{figure}

For a defect-free terrace of ideally parallel steps, the surface-state 
electrons on each terrace are confined perpendicular to steps but are 
free parallel to them. Thus, the wave functions of surface-state 
electrons are separable, with standing waves normal to the steps and 
plane waves parallel to the steps. This leads to a wave pattern {\em 
perpendicular} to the steps. Fig.\ \ref{large_terrace}a shows these 
well known spatial oscillation of the surface-state electrons 
{\em perpendicular} to pre-existing step edges on a large terrace of 51 nm 
in width. In addition, surface impurities act as point defects and 
lead to radial wave patterns centered on the impurities. 
At 32 mV the waves have the expected wavelength of 
$\lambda/2=3.5$ nm (See Ref.~[\onlinecite{buergi00}]). At 7 K five maxima are 
discernable already in the topographic image. Note the irregularities 
of the plane waves introduced by the natural deviation of the step 
edge position from a straight line.

The LDOS (as recorded in dI/dV spectra) 
displays a step-like function marking the sharp increase of 
differential conductivity around the surface-state band onset (Fig.\ 
\ref{large_terrace}b). Following Ref.~[\onlinecite{li98}], we measure 
the onset of the surface-state, $U_{on} = (U_{top}+U_{bot})/2$, and a 
width, $\Delta = U_{top}-U_{bot}$, of 
the onset by continuing the slope at the middle of the rise to the 
bottom and the top of the rise as indicated in Fig.\ 
\ref{large_terrace}b. The values of $U_{on}= -(68\pm4)$ meV and $\Delta 
=8$ meV are in good agreement with earlier measurements.\cite{buergi99}

Confinement on narrow terraces leads to both a shift in energy of the
surface-state related peak and a broadening of it.\cite{morgenstern02}
Moreover, additional maxima develop. The example shown in Fig.\
\ref{large_terrace}c, taken on a terrace of 6.1 nm width, displays
onsets at -53 meV and 32 meV with onset widths of $\Delta_{-53 meV} =
23$ meV and $\Delta_{32 meV} =32$ meV, respectively.  These two peaks
correspond to the first two transverse modes of the surface state
confined by the terrace edges.

\begin{figure} %fig2 
\includegraphics{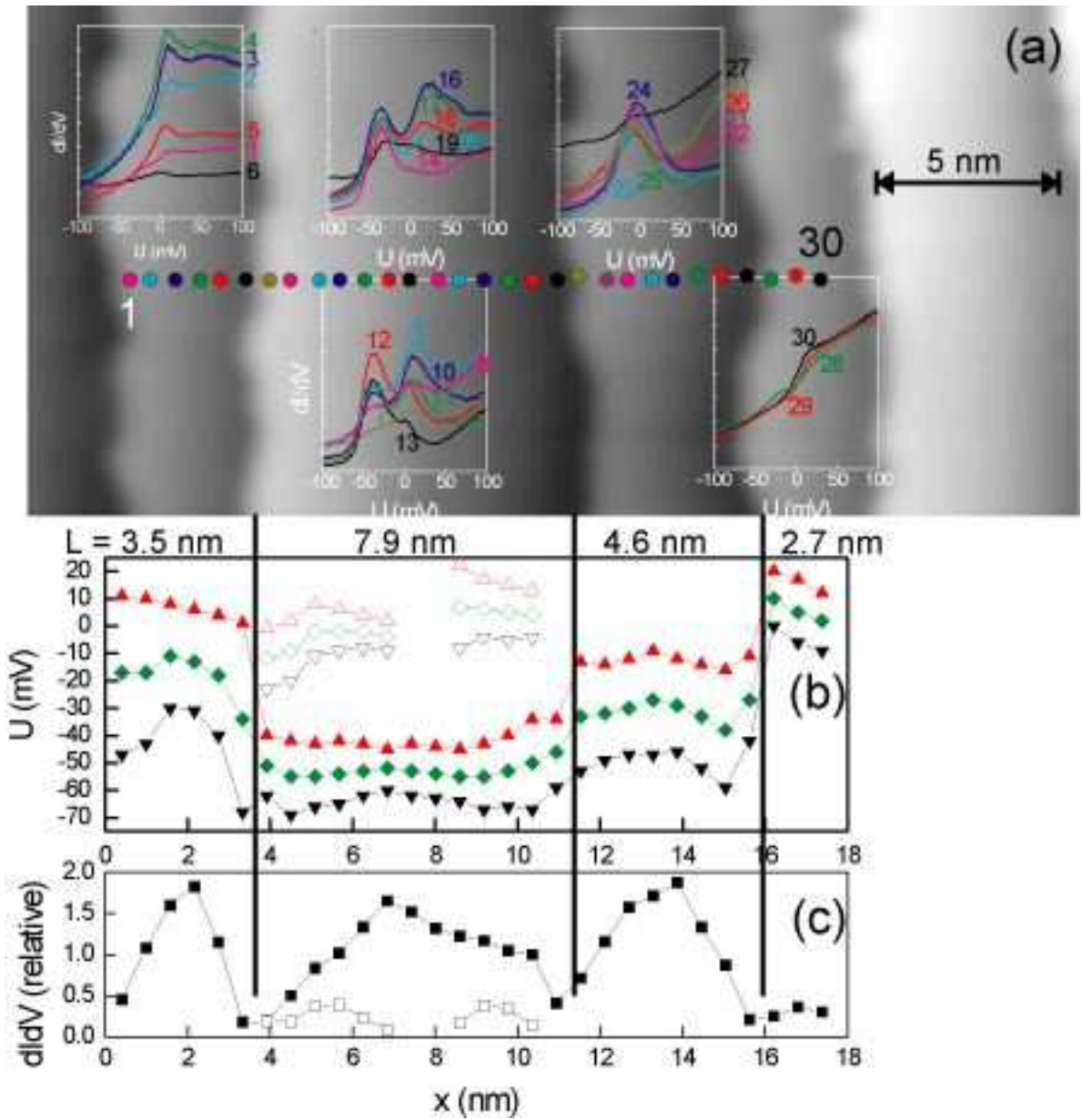} 
\caption[]{(Color online.) Position dependent STS at different points across the width of different terraces showing the dependence of the surface-state onset and of the second maximum in dI/dV 
spectra on average terrace width and on position on the terrace. (a) The background grayscale image is a topographic image ($U_{bias}$=-47 mV, $I_{tun}$=1.3 nA) showing approximately 5 terraces with irregular, disordered edges.  The average widths range from 2.7-7.9 nm. Superimposed on each terrace of the STM image is a set of inset STS data taken at various positions (indicated in color) across the terrace.  
The positions are numbered from left (1) to right (30).
(b) Position of bottom (down triangle), top (up triangle), and middle (diamond) of 
first (filled symbols) and second (open symbols) peak in dI/dV spectra of (a). The second peak is only observed on the widest terrace (of width 7.9 nm) and is related to exciting the second transverse mode. Dark vertical lines 
indicate positions of the step (terrace) edges.
(c) Relative intensity [i.e. ($dI/dV_{top}-dI/dV_{bot})/dI/dV_{bot}$] of the first peak 
(filled symbols) and the second peak (open symbols) for spectra of (a). 
\label{perp}} 
\end{figure}

The spectra vary for terraces of different width, but also for
different positions on the same terrace (Fig.\ \ref{perp}) with slight
variations in apparent surface state onset energy and large variations
in dI/dV intensity (Figs.\ \ref{perp}b and c). In particular, close to
the step edges, $U_{bot}$ shifts to lower values influencing the onset
value $U_{on}$ (Fig.\ \ref{perp}b). We have shown before that these
variations close to the step edges are in accordance with a
Fabry-P\'{e}rot model with the steps acting as finite,
energy-dependent barriers, so this is no indication of a shift in
energy.\cite{morgenstern02} As a consequence of strong scattering of
the surface-state electrons by the step potential, the amplitude of
the surface-state is smaller close to the step edges.\cite{everson90}
Thus, the peak intensity of dI/dV varies for the same terrace width
depending on the distance from the step edges (Fig.\ \ref{perp}c),
being highest in the middle between the step edges for the first peak
in dI/dV and showing two maxima for the second peak in dI/dV on the
largest terrace. Only the widest terrace, the second one in
Fig.~\ref{perp}a, shows two peaks in dI/dV.

\begin{figure} %fig2_2 
\includegraphics{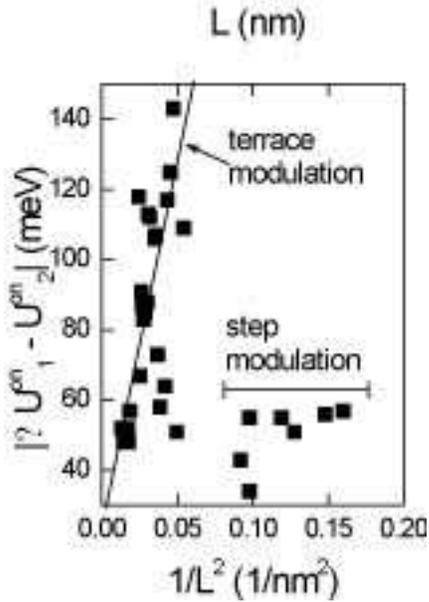} 
\caption[]{
Scaling of the relative position of the first and second maximum in dI/dV curves on small terraces vs. inverse terrace width squared. For terraces wider than $L\approx 4.3$ nm, $|U^{on}_1-U^{on}_2|\propto L^{-2}$ as shown by the linear fit, implying an interpretation consistent with the presence of transverse modes of the surface state. However, the effective mass determined, $m^{terr}_{eff} = (0.59\pm 0.17)$ $m_e$, deviates from the
effective mass of electrons in the surface-state band where
$m_{eff}=0.4 m_e$. On the other hand, terraces narrower than $L\approx 3.4$ nm do not scale linearly with $L^{-2}$, but instead show behavior consistent with step modulation.
\label{perp_2}} 
\end{figure}

The second peak arising in dI/dV due to local confinement on a single
terrace should show a quadratic dependence of energy with terrace
width according to a particle-in-a-box model for the simple case of
infinitely high potential walls:\cite{hansemann03}
$E_{gnd}=\hbar^2\pi^2/2m_{eff}L^2$, with $m_{eff}$ the effective mass
and $L$ the terrace (confinement) width.  The plot of the energy of
the second peak relative to apparent surface-state onset vs. $1/L^2$
(Fig.\ \ref{perp_2}) shows indeed a linear dependence for terrace
widths $>\lambda_F/2$, i.e.\ for those terraces where the
Fabry-P\'{e}rot model is applicable.\cite{morgenstern02} From the
prefactor of ($6.371 \pm 1.794)$ eV nm$^2$, we determine
$m^{terr}_{eff} = (0.59\pm 0.17)$ $m_e$, which deviates from the
effective mass of electrons in the surface-state band where
$m_{eff}=0.4 m_e$.

\begin{figure} %fig3 
\includegraphics{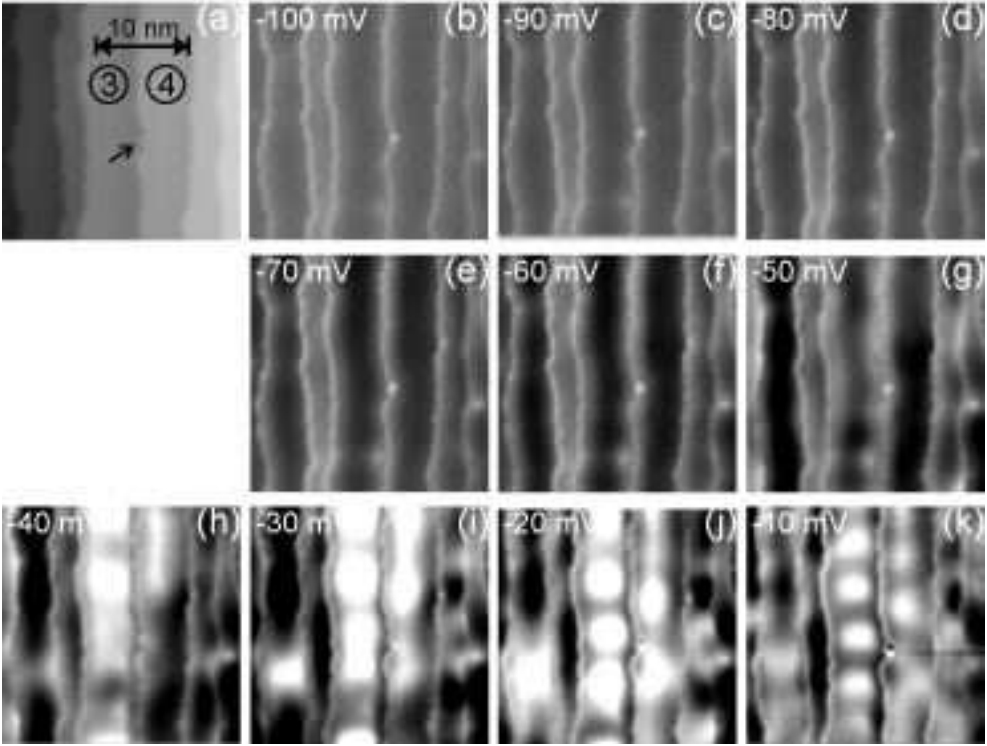} 
\caption[]{dI/dV maps of Ag(111) terraces in the occupied region of the spectrum [negative tip-sample voltage bias, indicated in the upper left-hand corner of (b)-(k)]. The AC voltage modulation is  $U_{mod} = 4$ mV and the frequency is $\nu_{mod}=738.1$ Hz. 
(a) Topographic image ($U_{bias}$=-100 mV, $I_{tun}$=13 nA). The two central terraces are labeled ``3'' and ``4''. The arrow points to an impurity at step a edge between the two. (b)-(k) Show dI/dV maps for $U_{bias}$ = -100 mV to -10 mV as indicated. Note the difference in wave patterns 
on the different terraces.  In particular, the terraces ``3'' and ``4'' begin to develop oscillations in the LDOS parallel to the step edges over the range of tip-sample bias $-40\; {\rm meV} \lesssim U_{bias} \lesssim -10$ meV.  The oscillations on terrace ``3'' are investigated in more detail in Fig.~\ref{compare_mapnegative}.
\label{didv_occupied}} 
\end{figure}

\begin{figure} %fig4 
\includegraphics{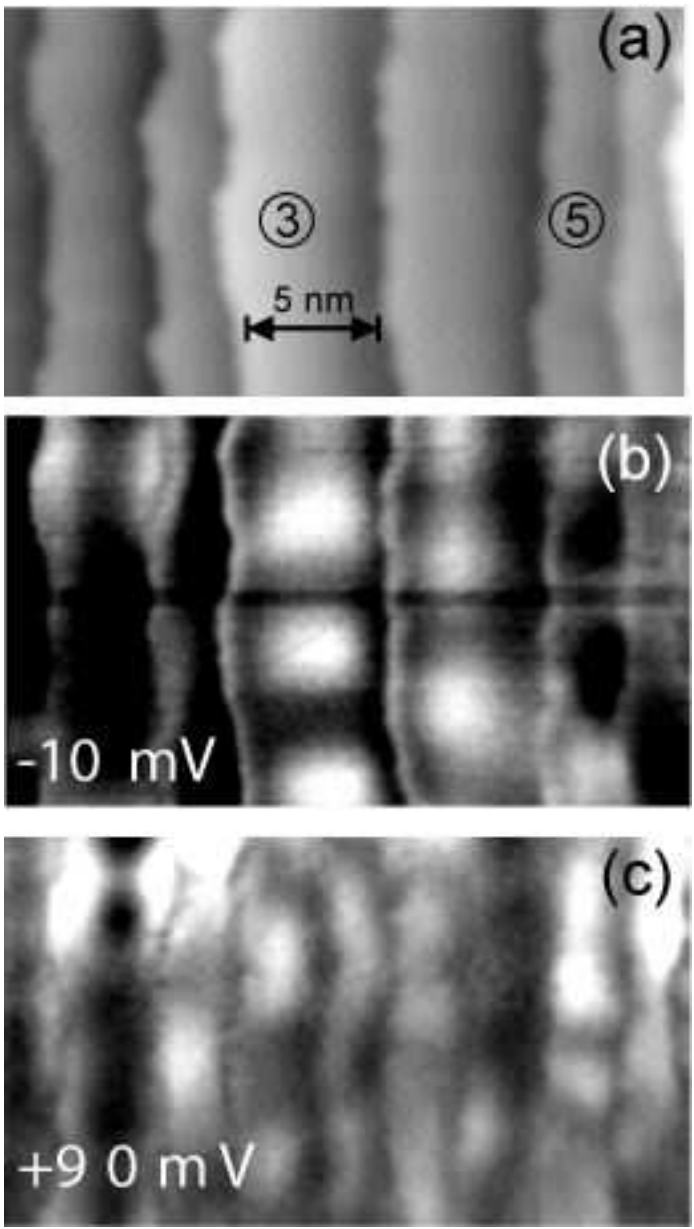} 
\caption[]{ 
Comparison of dI/dV maps for different tip-sample polarity with AC modulation parameters $U_{mod} = 4$ mV, $\nu_{mod}=738.1$ Hz and $I_{tun}=1.4$ nA. 
(a) Topographic image of approximately 6 irregular terraces.  The terraces are the same as those shown in Fig.~\ref{didv_occupied} (upper half).
(b) dI/dV map at $U_{bias}=-10$ mV. 
(c) dI/dV map at $U_{bias}= 90$ mV. 
Note the parallel and perpendicular wave pattern on terrace ``3'' for negative and positive polarity, respectively. This change in the wave pattern can be understood with the scattering theory described in the text. See Fig.~\ref{compare_mapnegative}.
\label{vergleich}} 
\end{figure}

\begin{figure} %fig5 
\includegraphics{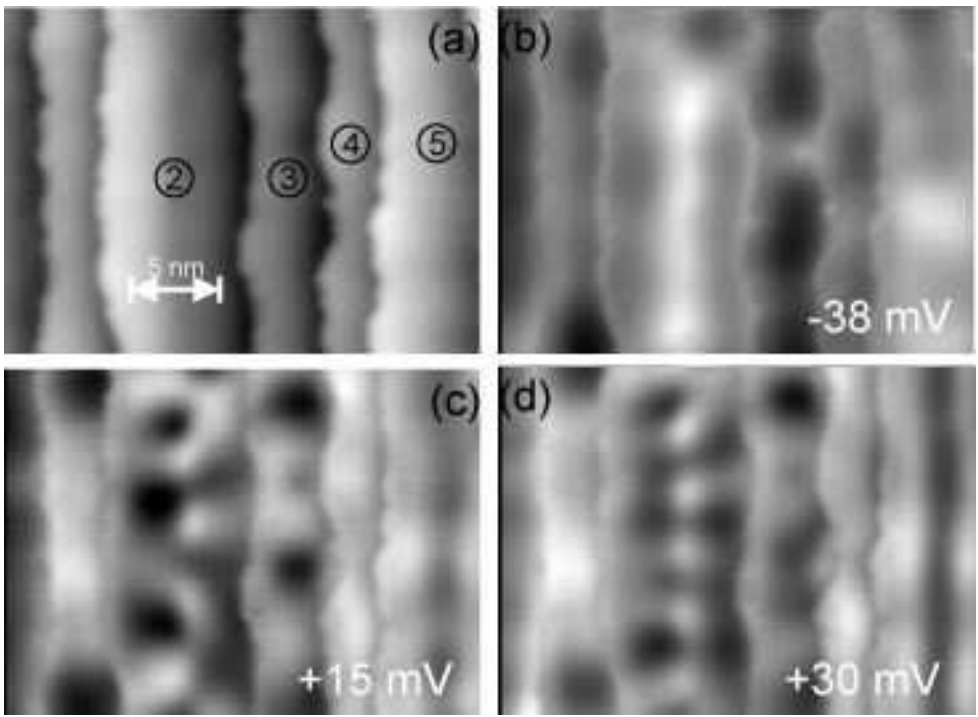} 
\caption[]{ 
Evolution of terrace wave patterns with $U_{bias}$.  The terraces shown are different from the set shown in Figs.~\ref{didv_occupied} and~\ref{vergleich}. The STM parameters are  $U_{mod} = 4$ mV, $\nu_{mod}=327.9$ Hz and $I_{tun}$=1 nA. 
(a) Topographic image taken at $U_{bias}=15$ mV, $I_{tun}$=1 nA. Terraces ``2'',``3'',``4'', and ``5'' develop distinct wave patterns with varying $U_{bias}$, indicating a
clear dependence of wave pattern on both terrace edge disorder and terrace width. 
(b)-(d) dI/dV maps at -38 mV, +15 mV, and +30 mV. Note the interesting superposition of 
parallel and perpendicular wave patterns on terrace ``2''. 
\label{vergleich_broad}} 
\end{figure}

\begin{figure} %fig6 
\includegraphics{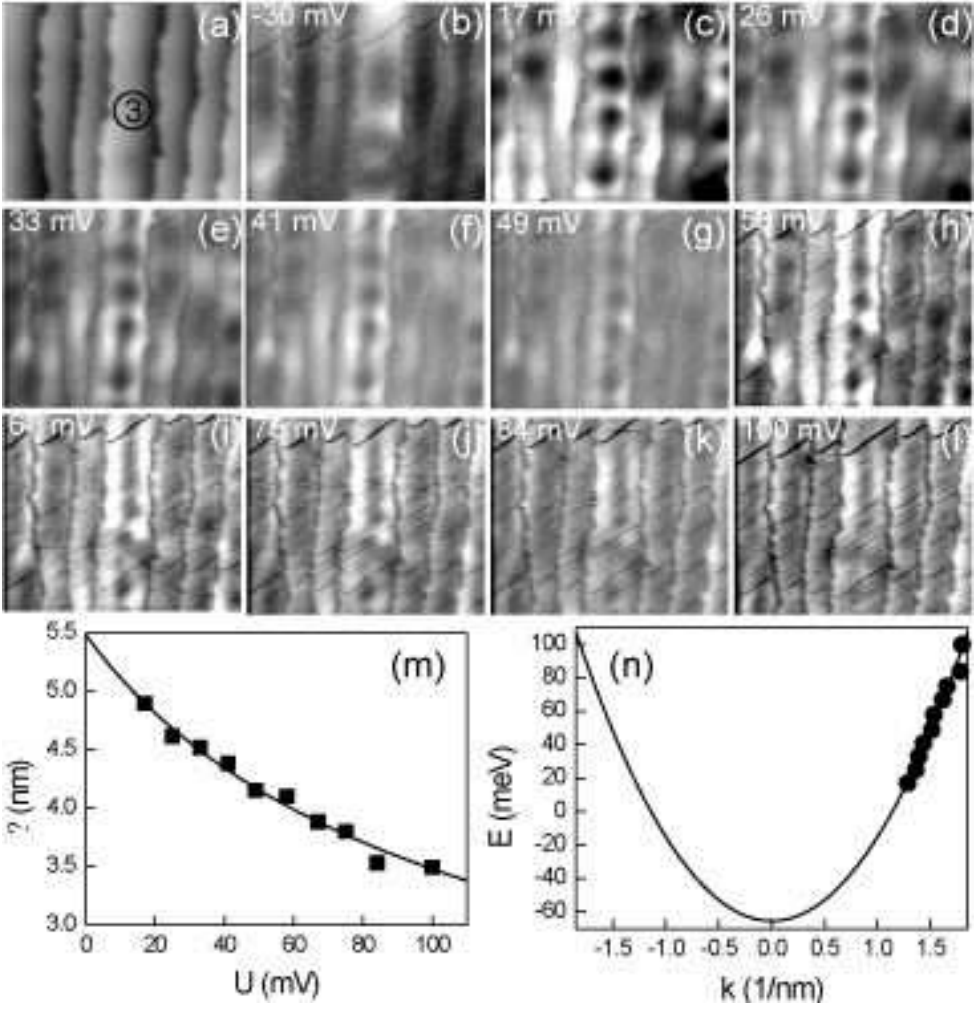} 
\caption[]{Wavelength dependence of parallel wave patterns in unoccupied region of the energy spectrum (positive tip-sample bias).  
Compare with Fig.~\ref{didv_occupied} which shows different terraces at negative $U_{bias}$. STM parameters are $U_{mod} = 8$ mV and $\nu_{mod}=381.8$ Hz. 
(a) Topographic image (-30 mV, 0.39 nA). 
(b-l) dI/dV maps for $U_{bias}$ = 17 mV to 100 mV and -30 mV as indicated. 
Black stripes (in h-l) are an artifact of the STM electronics and should be ignored. 
For better visibility (h)-(l) have four times the intensity contrast of (c)-(g). Note that there is a ``contrast'' reversal of the wave pattern of terrace ``3'' compared to those image shown in Fig.~\ref{didv_occupied}.  This feature is also captured by the scattering theory.  (See Fig.~\ref{compare_mappositive}.)
(m) Wavelength of standing waves on terrace ``3'' in (b) to (l) with square root fit. (See Sec. IIB in paragraph just below Eq.~(\ref{eq:E_n}).) 
(n) Dispersion relation as deduced from the fit in (m). 
\label{didv_unoccupied}} 
\end{figure}

To investigate spatial variations of the differential conductance in 
more detail, we have recorded dI/dV maps on these step arrays at 
different energies. Some examples are displayed in Figs. 
\ref{didv_occupied} to \ref{didv_unoccupied}.

Below the onset of the surface-state, dI/dV maps display bright lines 
and spots on a continuous background (Fig.\ \ref{didv_occupied}b to 
d). The bright lines correspond to the surface steps and are an 
artifact of the constant current scanning process (see above). The 
brightest spot in the dI/dV maps corresponds to an impurity at the 
step edge. It is visible in the topography as a protrusion
imaged 20 pm higher than the surrounding step edge 
(Fig.\ \ref{didv_occupied}a). 
This impurity can be used as a reference point to compensate for 
thermal drift. The two less bright spots are not visible in the topographic 
STM image in the contrast shown. Contrast enhancement shows them as 
protrusions of only 4 pm in height. We therefore attribute them to 
subsurface impurities.

With increasing voltage a difference in contrast develops between 
terraces of different width. 
For the broadest terrace ``3'', there is a contrast switch from 
dark to bright between -50 mV and -40 mV (Fig.\ \ref{didv_occupied}g to h). 
Narrower terraces switch contrast at voltages closer to $E_F$. 
The dependence of the switch in intensity on terrace width results from the 
different apparent surface-state onsets (see Fig.\ \ref{perp}a). 
On the largest terrace ``3'' a wave pattern evolves {\em parallel} 
to the step edges. The wavelength of this wave pattern increases as 
the energy is lowered with respect to $E_F$. While a faint wave 
pattern discernable at -50 mV on terrace ``3'' (Fig.\ 
\ref{didv_occupied}g) seems to originate from the subsurface impurity, 
the standing wave pattern at smaller energies is only perturbed by 
this impurity. The wave pattern does not originate from it.

There is no obvious correspondence between wave patterns on adjacent 
terraces. For example, at -30 meV (Fig.\ \ref{didv_occupied}i) the distance 
between the uppermost maxima is different on terrace ``3'' and on 
terrace ``4''. At -10 mV (Fig.\ \ref{didv_occupied}k) the wave patterns on 
terraces ``3'' and ``4'' are phase shifted in the upper half of the 
image.

Fig.\ \ref{vergleich} compares the parallel wave pattern (-10 mV, 
Fig.\ \ref{didv_occupied}k) to a pattern at opposite polarity, i.e.\ 
in the unoccupied region (+90mV). At +90 mV the dominant wave pattern 
on terrace ``3'' runs perpendicular to the step edges, however, with 
contrast variations. Thus, perpendicular and parallel wave patterns are 
superimposed. Note that even at +90 meV the narrower terraces, e.g.\ ``5'', show the 
parallel wave pattern only.

This superposition of parallel and perpendicular wave patterns becomes 
more pronounced on even broader terraces (Fig.\ 
\ref{vergleich_broad}). For the terrace ``2'' of 8.5 nm in width, the 
perpendicular wave pattern dominates at -38 mV (Fig.\ 
\ref{vergleich_broad}b). At +15 mV and +30 mV the contrast variation 
along the perpendicular waves on the left half and on the right half 
of the terrace are independent of each other. This leads to an 
interference pattern at the middle of the terrace, which changes 
wavelength with energy (Fig.\ \ref{vergleich_broad}c to d). Note the 
differences of terrace ``3'' from the slightly narrower terrace ``5'' at 
+ 15 mV and +30 mV. Terrace ``5'' shows lower conductivity at its 
center than at its sides, while terrace ``2'' shows higher 
conductivity at its center compared to its sides. Also terraces ``3'' and 
``4'' show wave patterns of opposite contrast.

The wavelength dependence on energy of the parallel wave pattern is 
investigated in more detail on the unoccupied side of the LDOS (Fig.\ 
\ref{didv_unoccupied}). The standing wave pattern {\em parallel} to 
the surface steps on terrace ``3'' at positive polarity is 
contrast-reversed with respect to the standing wave pattern shown in 
Fig.\ \ref{didv_occupied}. The dI/dV map at -30 mV (Fig.\ 
\ref{didv_unoccupied}b) shows that this is an effect of the opposite 
polarity of the tip-sample voltage bias and not of the terrace width. Due 
to a subsurface impurity (see below) the wave pattern in the lower 
part of this image is less regular. Thus, we concentrate on the upper 
part of the image.

The average wavelength changes with bias voltage (Fig.\ 
\ref{didv_unoccupied}m). For an ideal terrace with impenetrable 
edges, the isotropic free surface dispersion relation $E({\vec 
k})=E_0+\frac{\hbar^2 k^2}{2m_{eff}}$, with $E_0=-0.065$ eV, 
is modified by ``confinement'' due 
to the terrace structure, 
\begin{equation} 
\label{eq:E_n} 
E_n(k_\|)=E_0+\frac{\hbar^2 \pi^2 n^2}{2 m_{eff} L^2} 
+\frac{\hbar^2 k_\|^2}{2m_{eff}}\;, 
\end{equation} where $k_\| \equiv 2\pi/\lambda_\|$ and $n$ is the 
number of transverse modes. This equation predicts that for a fixed 
energy, narrower terraces with smaller width $L$ will have a larger parallel 
wavelength, $\lambda_\|$. Examples of this effect can be seen in 
Figs.\ \ref{didv_occupied} to \ref{didv_unoccupied}, although the 
oscillations in the narrower terraces are less regular for the 
narrower terrace. We attribute this to larger relative variations in 
width for fixed edge variations in the narrower terraces.

Furthermore, a quadratic variation of $\lambda_\|$ with $E$ is 
expected for a free electron dispersion (see Eq.\ (\ref{eq:E_n})) 
provided no additional transverse modes are occupied over the energy 
range probed. Indeed, the data can be well fitted by 
$a(E-E_0)^{-1/2}$. We find $E_0 = -(67\pm 5)$ meV, the 
same value as the bottom of the surface-state band determined from the 
surface-state onset on large terraces (see above) and from 
perpendicular wave patterns on large terraces. \cite{jeandupeux99} 
Also the wavelengths determined from the negative dI/dV maps of Fig. 
\ref{didv_occupied} fit a quadratic dispersion with $E_0 = -(68\pm 
19)$ meV. Fig.\ \ref{didv_unoccupied}n displays the dispersion 
relation deduced from this measurement that reproduces the dispersion 
relation determined from perpendicular wave patterns. \cite{buergi00} 
Thus, parallel wave patterns originate also from the scattering of 
surface-state electrons.

\section{Theory} 
\label{sec:theory}

\subsection{Model and approach}

In this paper we are focused on understanding the role of disorder in 
the terrace edge profile on the observed modulations in the local 
density of states. Disorder in the edge profile breaks translational 
symmetry along the direction parallel to the terrace edges. This 
broken symmetry prevents a simple separation of the energy (and the 
wavefunctions) of the electrons into a parallel and transverse 
component, as was done in Eq.~(\ref{eq:E_n}). In order to conveniently 
handle the broken translational symmetry another approach is 
needed. We chose to use a scattering theory\cite{fiete03} that has been quite 
successful in predicting the standing wave patterns in the LDOS observed with 
the STM around impurities and step edges on the surfaces of the noble 
metals that support surface-states.

The scattering theory given in Ref.~[\onlinecite{fiete03}] most
naturally describes the scattering of electrons in the surface-state
from point-like impurities (adatoms) where an s-wave scattering
approximation can be made; scattering from an extended impurity, such
as a terrace edge, requires some modification of the theory, which
otherwise cannot directly describe scattering from an extended object.
Fortunately, since the standing wave patterns we are interested in
(those parallel to the terrace edges) are observed most clearly in the
central portion of the terrace, we may make a simple approximation in
the scattering theory for the terrace edge: We treat the terrace edge
as a dense ``row'' of point (s-wave) scatterers. Far from the terrace
edges (near the middle of the terrace, for example) the results will
be indistinguishable from a terrace edge that is continuous. A ``row''
of atoms is expected to be a good approximation to the terrace edge
provided the atoms that make up the wall are well within a wavelength
(the wavelength of the surface-state electron) of each other and the
distances probed are of the order of the separation of the atoms or
larger.  In the calculations described below our objective will be
first to understand the qualitative features we can expect from
scattering theory and second to compare as directly as possible with
experiment.  The latter requires determining the phase shift of a
single atom (one of many that make up the terrace edge) so that the
correct phase shift results for the edge as a whole. Since the input
to the scattering theory is the wall profile, the dispersion of the
surface-state electrons, and the s-wave scattering phase shift
$\delta$, the only free parameter we have in the theory is
$\delta$. We tune $\delta$ for a fixed wall profile and surface-state
dispersion to reproduce the experimental dI/dV vs. V spectra at the
center of the terrace. Once $\delta$ is optimized in this way, we
computed the dI/dV image maps (at a fixed voltage) over all positions
using this value and the exact edge profile of the terraces.

\subsection{Theoretical calculations}

\subsubsection{Model calculation for qualitative effects}

\begin{figure} %fig7 
\includegraphics{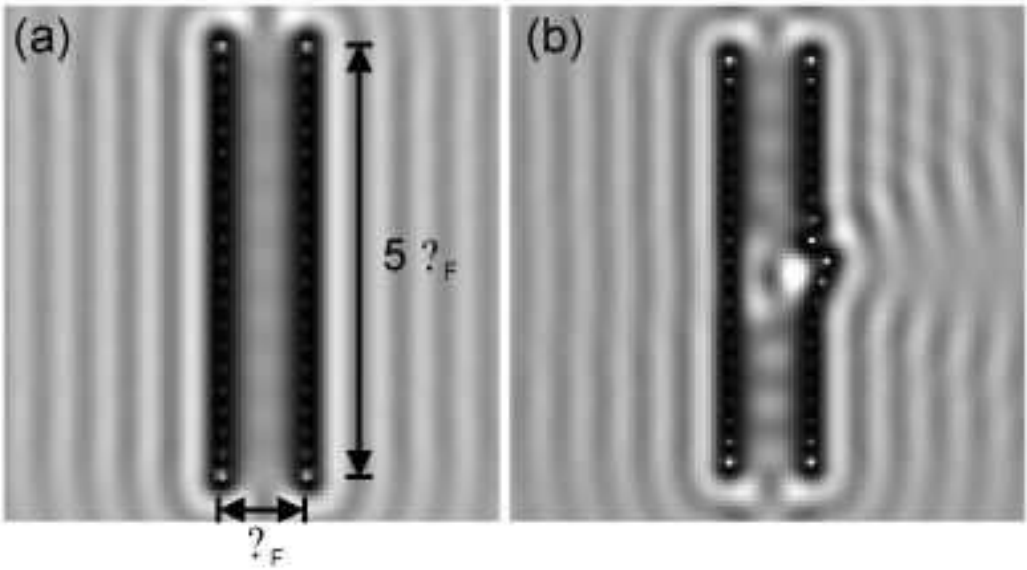} 
\caption[]{Model scattering theory calculations illustrating creation of standing wave patterns in the LDOS parallel to the terrace edges when disorder is present.  {\bf Here ``?''=``$\lambda$'' due to a graphics error.} The terrace edges are modeled by rows of atoms.  As long as the atoms are much closer together than the wavelength of the electrons in the surface state this is a good approximation.  Shown are two $\left[dI/dV\right]_{\rm calc}$ maps calculated with scattering theory for a terrace of length 
$5\lambda_F$ and width $\lambda_F/4$ with
(a) no defect and 
(b) one defect. 
The introduction of a single point defect creates a strong variation in the LDOS parallel to the 
edges of the terrace. 
In both cases there are oscillations clearly visible perpendicular to the terraces edges outside the terrace. 
%Note that the x and y-scales are different.  This is an artifact of the plotting program used to generate the figure, and does not affect the physical point being made.
\label{theorie_unitless}} 
\end{figure}

To understand the qualitative effects we expect from a scattering 
theory approach to the LDOS modulations from disordered 
terraces, we first carried out calculations on a single terrace with 
perfectly straight walls and then added in increasing amounts of 
disorder. The terraces were modeled by placing 21 point (s-wave) 
scatterers at a distance of $\lambda_F/4$ along the terrace edge. 
See Fig.~\ref{theorie_unitless}.

The dI/dV maps were computed using the surface-state dispersion $E({\vec 
k})=E_0 + \frac{\hbar^2 k^2}{2m_{eff}}$. (Confinement effects that do 
not cause an apparent shift in the surface-state band minimum are 
captured automatically by electron scattering from the terrace 
``edges''.) An approximation to the ``constant current'' condition 
used in the experiment (while simultaneously aquiring topographic and 
dI/dV maps) is achieved by calculating 
\begin{equation} 
\label{eq:dIdV} 
\left[\frac{dI}{dV}(eV)\right]_{\rm 
meas}\equiv\left[\frac{dI}{dV}(eV)\right]_{\rm calc}- \langle 
\left[\frac{dI}{dV}\right]_{\rm calc} \rangle 
\end{equation} 
where $\langle 
\left[\frac{dI}{dV}\right]_{\rm calc} \rangle = 
\int_0^{eV} d\epsilon \left[\frac{dI}{dV}(\epsilon)\right]_{\rm calc}(eV)^{-1}$. This procedure correctly reproduces brightness variations in dI/dV maps 
simultaneously taken with topographic maps.\cite{fiete01} 

\begin{figure} %fig8 
\includegraphics{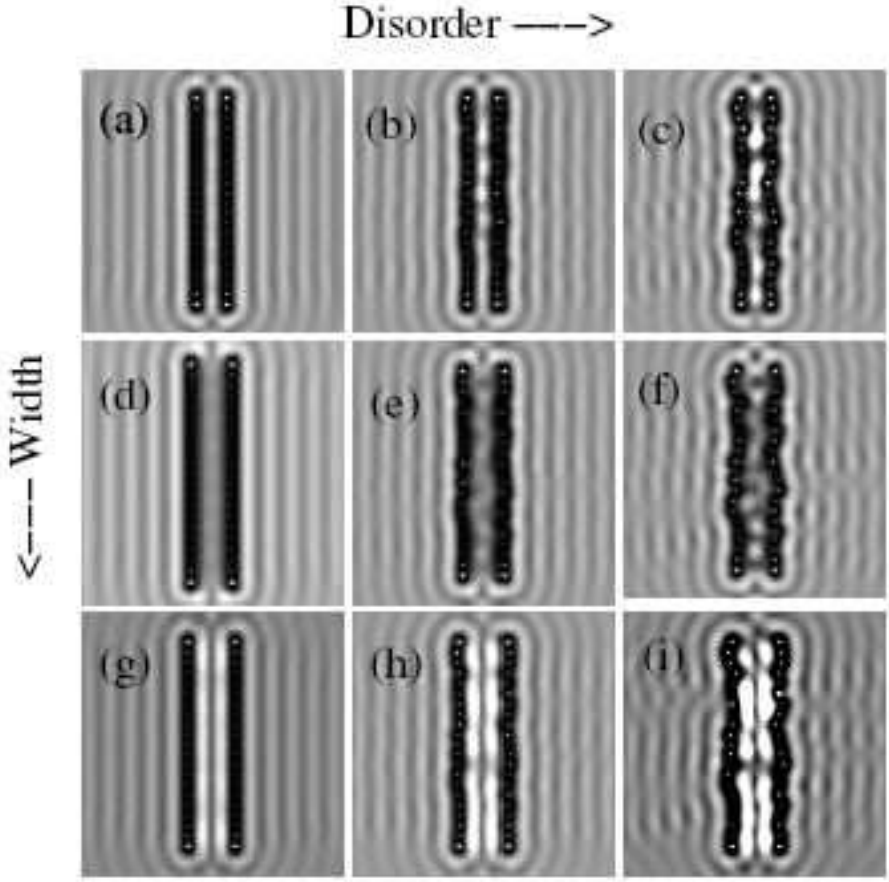} 
\caption[]{ 
Theoretical dI/dV maps for different width and disorder. 
Disorder is introduced manually by shifting the horizontal positions of the atoms that model the walls of the terrace.  The characteristics of the disorder (such as whether or not the displacements are Guassian distributed, what the RMS value of the fluctuations is, etc.) are unimportant for the qualitative study here. The maximum displacement of the ``random'' positions is:  (a,d,g) 0, i.e., no disorder, 
(b,e,h) $\pm 8$\% variation in width, and 
(c,f,i) $\pm 20$\% variation in width. The average widths of the terraces
are:
(a,b,c) $0.8 \lambda_F$, 
(d,e,f) $\lambda_F$, and 
(g,h,i) $1.2 \lambda_F$.  The features seen in (c) and (i) are qualitatively similar to those seen on terrace ``3'' in Fig.~\ref{didv_occupied} at negative and positive voltage biases, respectively. See Fig.~\ref{compare_mapnegative}.
\label{theorie_9images}} 
\end{figure}

In these initial model calculations, each scatterer is assumed to have
a phase shift of $\pi/2$. This value turns out to be almost ``universal''
for adatoms on the surfaces of noble metals\cite{fiete03} so we take
it as a starting point; later we will find a different value matches
the experiments better when we attempt to make a direct, quantitative
comparison between theory and experiment. The initial choice of
$\pi/2$ is arbitrary and does not affect any of the qualitative
conclusions reached from the calculations presented in
Figs.~\ref{theorie_unitless} and~\ref{theorie_9images}.  While this
value works well for single adatoms, the physics of scattering from a
terrace edge is very different from the physics of scattering from an
isolated atom, and this is why the best phase shift for the terrace
edge atom differs from $\pi/2$.

Fig.\ \ref{theorie_unitless}a shows the calculated 
$\left[dI/dV\right]_{\rm calc}$ map for a trough of $5 \lambda_F$ in 
length and $\lambda_F/4$ in width limited by 21 point scatterers on 
each side. While the perpendicular standing wave pattern outside the 
trough is clearly visible, its narrowness inhibits a similar wave 
pattern within the trough. It is too small to support even half of a 
Fermi-wavelength. The introduction of an arbitrary defect in one of the 
terrace walls leads to standing waves parallel to the rows of atoms 
with the same wavelength $\lambda_F/2$ as the perpendicular wave 
outside the trough (Fig.\ \ref{theorie_unitless}b). Note the faint 
signs of oscillations in the ``perfect'' terrace as well. These 
originate from the finite length of the trough, where its ends act as 
defects. These first calculations demonstrate that a parallel 
wave pattern is indeed a result of terrace disorder.

For a better understanding of particular features in the wave patterns
observed in the experiments, we compare $\left[dI/dV\right]_{\rm
calc}$ maps of varying terrace width and terrace disorder (Fig.\
\ref{theorie_9images}). Terrace disorder has been modeled in the way
described in the figure caption. Fig.\ \ref{theorie_9images} a, d, and
g support the idea that a wave pattern perpendicular to the steps can
only develop for larger terraces. The intensity of the parallel wave
pattern increases with increasing disorder (Figs.\
\ref{theorie_9images}a to c) and shows the same qualitative features
as in the experiments. For example, some of the maxima in Fig.\
\ref{theorie_9images}c are brighter than neighboring ones; some of
them seem to be disconnected from neighboring maxima by deep minima,
while between other maxima there are only little contrast
variations. Fig.\ \ref{theorie_9images} c and i show that two bright
protruding lines are seen instead of one for a larger terrace width.
For the broadest disordered terrace (Fig.\ \ref{theorie_9images} h, i)
the two wave patterns are thus superimposed. Minima in the two
perpendicular patterns are displaced with respect to each other by up
to half a wave length (Fig.\ \ref{theorie_9images}i) as in the
experiment. Many of the qualitative features of the experiments can
thus be reproduced by the scattering theory, showing they result from
terrace edge disorder.

\subsubsection{Direct comparison with experiment}

\begin{figure} %fig9 
\includegraphics{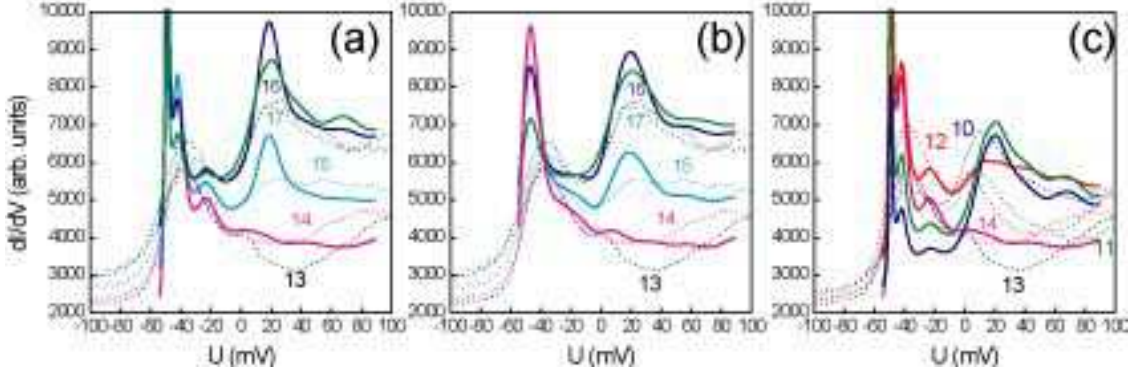} 
\caption[]{(Color online.) 
Comparison of dI/dV spectra taken at different positions in the second terrace in Fig.~\ref{perp}a of width $7.9$ nm. Solid lines: calculation with phase shift $0.9\pi$ in the surface-state scattering theory described in the text. 
Dotted lines: Experimental results.
(a,c) Scattering calculation without thermal broadening.
(b) same as (a), but theoretical curve is averaged over an energy window to simulate broadening due to thermal effects and modulation voltage.  The quality of the agreement between experiment and theory in (b) is good considering the range of energy and the relative simplicity of the theory.  The main disagreement occurs around $U\approx -50$ meV where the theory predicts a narrower and higher peak than is seen experimentally.  This indicates the approximations used to model the terrace edge result in a more highly reflective edge at these energies than the actual edge in experiment.
\label{compare_dIdV}} 
\end{figure}

The calculations presented so far were meant to illustrate that the
main physics is captured by the scattering theory and to emphasize the
quality of the qualitative agreement between the scattering theory and
experiment. Now, we go beyond these simple qualitative model
calculations and use the actual surface-state dispersion for Ag(111)
listed above and a terrace edge profile taken from experiment. The
surface-state onset on a large terrace lies at $E_0 = -0.065$ eV; the
effective mass is $m_{eff} = 0.4 m_e$.  These values lead to
$\lambda_F=7.38$ nm for the Fermi-wavelength on an infinite
terrace. First, we calculate position dependent
$\left[dI/dV\right]_{\rm calc}$ spectra for a 7.9 nm wide terrace with
varying phase shift and compare them to the experimental spectra for
the broadest terrace in Fig.\ \ref{perp}. The phase shift of the
scatterers is chosen to best reproduce the Fabry-Perot type
oscillations with the tip-sample bias, as shown in
Fig.~\ref{compare_dIdV}. The best agreement is reached with a phase
shift of $\delta = (0.9 \pm 0.1)\pi$. Those spectra resemble the
experimental ones with a shift in the apparent surface-state onset and
a second peak in the unoccupied region. Note that in order to keep our
fitting parameter to only the number $\delta$, we do not include any
energy dependence in the phase shift.

There is good agreement between the spectra calculated for the right
half of the terrace (Fig.\ \ref{compare_dIdV}a), if spectrum number 14
showing no second maximum is taken as the central spectra, though it
is measured 0.6 nm to the right of the geometrical center.  This might
be simply due to difficulties in determining the geometrical center of
the terrace edges or may be due to the asymmetry of the reflection
properties of ascending and descending step edges.\cite{jeandupeux99}
The reflectivities of steps on this surface have been measured to be
$r_{desc} =0.72 - 2.91 E/eV$ and $r_{asc}=0.48-2.86E/eV$
\cite{jeandupeux99} for the descending and the ascending step,
respectively.  Spectra 15, 16, and 17 are measured at a distance of 0.6
nm, 1.2 nm, and 1.8 nm from spectra 14 are compared to calculations
performed at `0.7 nm', `1.5 nm', and `2.0 nm' from the center of the
terrace, respectively. In all cases, the peak at $\approx 20$ meV lies
at almost exactly the same energy and its relative intensity is well
reproduced. The first peak lies also at approximately the correct
energy. However, a sharp rise in the theoretical spectra (an artifact
of the calculation procedure) makes a comparison difficult for the
first peak. In both theory and experiment, the second peak is broader
than the first peak, because the reflectivity and thus the lifetime of
the resonance decreases with increasing energy.

In Fig.\ \ref{compare_dIdV} the peaks are sharper and more intense in
theory than in experiment.  Several processes may explain this
difference in sharpness and intensity. On the experimental side, the
spectra show thermal and modulation voltage broadening. We simulate
the modulation voltage averaging and temperature broadening by
averaging the theoretical spectra (Fig.\ \ref{compare_dIdV}b). Indeed,
this broadening makes the additional peak in the theory at $\approx
-20$ meV undetectable in the experiment and lowers the intensity of
the peak at +20 meV.  However, the theoretical peaks remain somewhat
sharper, presumably from details associated with how the terrace edges
are modeled. One may think that the ``porous'' nature of the walls my
lead to a shorter lifetime, and hence a broader thoery peak than what
is observed experimentally.  However, this appears not to be the case.
Changing the number of atoms that make up the edge profile by 50\% has
a negligible effect on the widths of the peaks.

Discrepancies may result from not correctly modeling the energy 
dependence of the reflectivity by replacing the steps by point 
scatters: The peaks in dI/dV can be associated with Fabry-P\'{e}rot 
type resonance with an energy dependent 
reflectivity.\cite{morgenstern02} As a narrower resonance means a 
longer lifetime, and thus a greater reflectivity, we conclude that 
theory has an effective reflectivity which is larger than 
experiment. The calculation has an energy dependent reflectivity, but 
because the walls are built up from point scatterers, it is difficult 
to determine the reflectivity exactly in order to compare it directly 
with experimental results for a ``perfect'' terrace. With effort, a 
very close match could be made by making the phase shift energy 
dependent and possibly adding imaginary parts as well, 
but extremely accurate numerical agreement is not our goal 
here. Rather, we wish to understand what is responsible for the 
essential features of the data. A more accurate calculation than ours 
would also include changes to the surface-state band dispersion 
itself due to the terrace edges. Recall that our theory uses the 
surface-state dispersion of the free surface-states.

The dI/dV spectra to the left and to the right of the center differ in
both experiment and theory as a result of an asymetry in the disorder
of the two terrace edges. However, for the experimental spectra to the
left of the center, the second peak shifts with position on the
terrace. This is not recovered in the calculation (Fig.\
\ref{compare_dIdV}c). We attribute this difference to the fact that
the spectra are calculated for a terrace that is some 10\% smaller
than the experimental terrace and that has a different shape in
detail, on the one hand, and to the larger difference in reflectivity
properties between a row of atoms and a descending step than a row of
atoms and an ascending step, on the other hand.  The asymmetry in
scattering properties of the two steps that can not be captured by
replacing them by two rows of equivalent atoms leads thus to the
`off-center middle' (noted earlier in this section) and to larger
discrepancies on the half of the terrace that is closer to the
descending step edge. Despite these deviations in detail, the
scattering theory does satisfactorily reproduce the experimental
spectra. Thus, we have calculated dI/dV maps for a particular terrace
for a direct quantitative comparison of experiments to theory. We have
chosen terrace ``3'' of Fig.\ \ref{didv_occupied} due to the clear
wave pattern observed. This terrace is also broad enough that
changes to the surface-state on narrow terraces \cite{morgenstern02}
are negligible.

\begin{figure} %fig10 
\includegraphics{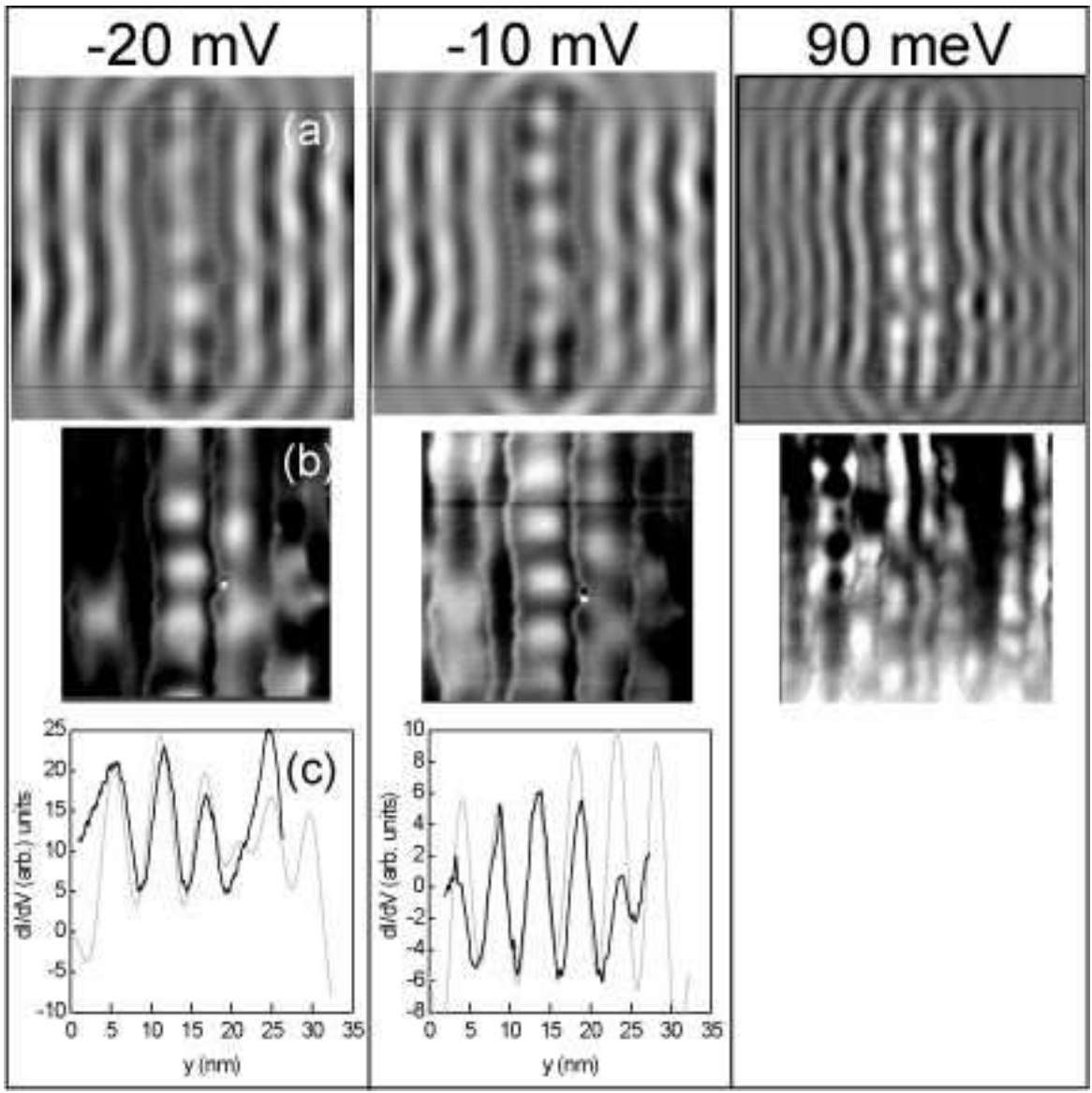} 
\caption[]{ 
Comparison of theoretical to experimental dI/dV maps for terrace ``3'' of 
Fig.\ \ref{didv_occupied} for bias voltages as indicated above the images. 
(a) Theoretical dI/dV maps computed using Eq.~(\ref{eq:dIdV}).
(b) Experimental dI/dV maps. Note that the field of view in y-direction is larger in the 
theoretical maps than in the experimental maps as indicated in (a) by the faint horizontal 
lines near the top and bottom of the figures.  The region in between these horizontal lines should be compared with the experimental figures in (b).
(c) We show line scans through the middle of terrace ``3''. Gray: theory.  Black: experiment. The 
y-axis is different in the two spectra for better comparison. 
\label{compare_mapnegative}} 
\end{figure}

The terrace walls are approximated by 54 atoms placed as close to the 
actual boundary as possible (Fig.\ \ref{compare_mapnegative}a, white 
dots). Fig.\ \ref{compare_mapnegative} compares the 
$\left[dI/dV\right]_{\rm meas}$ maps, computed from 
Eq.~(\ref{eq:dIdV}), of Fig.\ \ref{didv_occupied}j and k at -20 mV, at 
-10 mV, and of Fig.\ \ref{vergleich}c at +90 mV to the 
calculation. The qualitative agreement is excellent: A similar wave 
pattern is observed in the calculated spectra as in the experimental 
one, both in the occupied and the unoccupied region. Also the decrease 
in wavelength from -20 meV to -10 meV is reproduced in the 
calculation. The comparison of the $\left[dI/dV\right]_{\rm meas}$ 
line scans through the center of the terraces (Fig.\ 
\ref{compare_mapnegative}c) shows this good wavelength agreement for 
both -20 meV and -10 meV. The fourth maximum at -20 mV in Fig.\ 
\ref{compare_mapnegative}c is simply not resolved in the experimental 
data. Only the relative intensities are partly recovered.

\begin{figure} %fig11 
\includegraphics{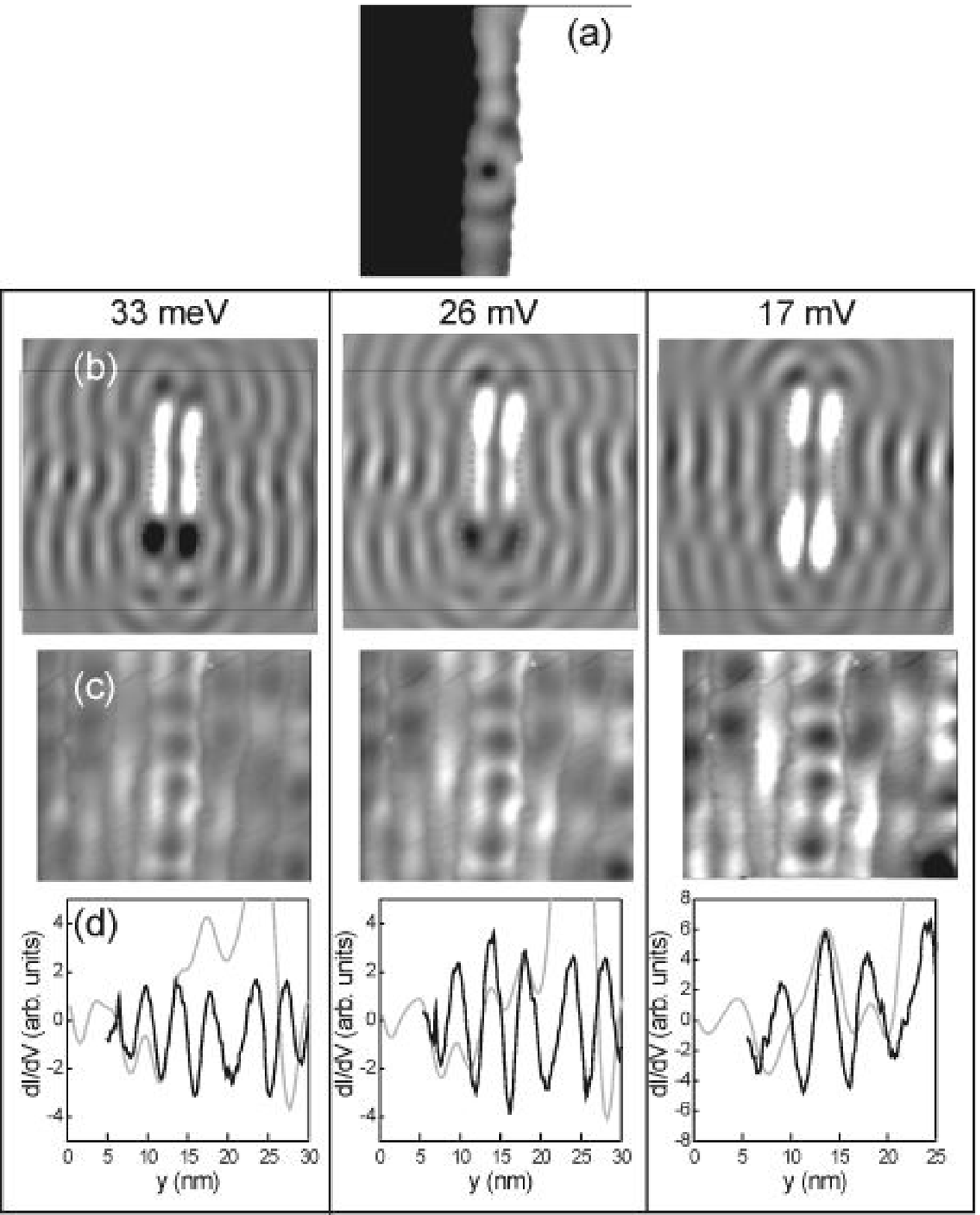} 
\caption[]{ 
Comparison of theoretical to experimental dI/dV map for unoccupied states (positive tip-sample bias).   
(a) Topography with enhanced contrast. Black dot on terrace indicates a point impurity. 
(b) Calculated dI/dV maps.  As in Fig.~\ref{compare_mapnegative} the region between the two faint horizontal lines should be compared with the experimental data in (c).
(c) Experimental dI/dV maps of Fig.~\ref{didv_unoccupied}. 
(d) Line scans through middle of terrace.  Gray: theory, Black: experiment. 
\label{compare_mappositive}} 
\end{figure}

We have also simulated the broadest terrace ``3'' in Fig.\ 
\ref{didv_unoccupied} for a more detailed comparison of theory with 
experiment for wavelength changes in the unoccupied region. Fig.\ \ref{compare_mappositive} 
shows the comparison of simulation and experiment at 33, 26, and 17 
mV. Unfortunately, the direct comparison is somewhat obstructed by an 
impurity, visible by enhancing the contrast in the 
topographic image (Fig.\ \ref{compare_mappositive}a) leading to a very 
particular wave pattern at -30 meV (Fig.\ \ref{didv_unoccupied}b). At 
first sight, the comparison seems to be less favourable. The 
simulation shows two bright protruding lines close to the step edges 
separated by a dark stripe with little contrast variations. However, 
this reproduces the variation of intensity across the terrace in the 
experiments (from bright to weaker to bright). Despite the 
impurity the lines cans through the middle of the terrace agree well in 
wavelength in the upper part of the image (Fig.\ 
\ref{compare_mappositive}d). Only a phase shift in the lower part is 
not recovered. 
Thus, again the major difference lies in intensity. 
This last example underlines the complexity of the evolved wave pattern 
and that it may depend quite sensitively on various details of the geometry.

\section{Conclusions} 
\label{sec:conclusions}

We have probed the LDOS of disordered terraces of varying width and
observed oscillations {\em parallel} to the terraces with a wavelength
in accordance with a free electron dispersion, and we have shown that
these parallel oscillations arise naturally in scattering theory
calculations for disordered terraces. The theory also captures
important contrast variations with tip-sample voltage polarity. Thus,
the wave patterns observed experimentally are a result of step
disorder.

The complete pattern is a combination of Fabry-P\'{e}rot (surface-state) and 
disorder scattering effects along with surface 
reconstruction effects. The Fabry-P\'{e}rot effects are intrinsic in 
scattering theory, but the changes to the underlying surface-state 
band structure are not (in the theortical model used here, 
which assumes a clean surface-state dispersion). As the scattering 
theory recovers all the qualitative features of the experiment and 
yields semi-quantitative agreement, the surface-state scattering 
effects dominate (at least for the wider terraces). Remaining 
differences are attributed to surface reconstruction effects. The 
semi-quantitative agreement in the simulation also shows that the 
cross-talk between the large terraces and their neighbors is 
negligible.

The terraces that the dI/dV maps have been calculated for are wide 
enough to allow us to use the surface-state band structure valid for 
infinitely broad terraces. In principle narrower terraces could be 
calculated within the same theory by using the appropriate dispersion 
relation for narrower terraces, which will in general be anisotropic and 
depend on terrace width. \cite{baumberger04}

As a future project, we suggest to locally investigate the reaction 
dependence on the vicinal surface and we expect local variation on the 
nanometer scale due to the reported variation of the LDOS.

\section{Acknowledgement} 
We acknowledge financial support by the Deutsche 
Forschungsgemeinschaft, NSF-9907949, NSF DMR-0227743 and the Packard Foundation. We 
are greatful for experimental support by K.-F.\ Braun, Ohio 
University, Athens. 
 
\newpage 
%%FIGURES%%%%%%%%%%%%%%%%%%%%%%%%%%%%%

\begin{thebibliography}{99} 
\bibitem{shockley39} W.\ Shockley, Phys.\ Rev 56, 317 (1939). 
\bibitem{forstmann70} F.\ Forstmann, Z.\ Phys.\ 235, 69 (1970). 
\bibitem{kevan89} S.D.\ Kevan, R.H.\ Gaylord, Phys.\ Rev.\ B 36, 5809 (1987)). 
\bibitem{roos89} P.\ Roos, E.\ Bertel, K.D.\ Rendulic, Chem.\ Phys.\ Lett.\ 232, 537 (1989). 
\bibitem{gartland75} P.O.\ Gartland, B.J.\ Slagsvold, Phys.\ Rev.\ B12, 4047 (1975). 
\bibitem{heimann77} P.\ Heimann, H.\ Neddermeyer, H.F.\ Roloff, J.\ Phys.\ C10, L17 (1977). 
\bibitem{zangwill88} A.\ Zangwill, Physics at Surfaces, Ch. 4 (Cambridge Univ. Press, Cambridge, 1988). 
\bibitem{tersoff85} J.\ Tersoff, D.R.\ Hamann, Phys.\ Rev.\ B31, 805 (1985). 
\bibitem{crommie93_nature} M.F.\ Crommie, C.P.\ Lutz, D.M.\ Eigler, Nature (London) 363, 524 (1993). 
\bibitem{crommie93_science} M.F.\ Crommie, C.P.\ Lutz, D.M.\ Eigler, Science 262, 218 (1993). 
\bibitem{avouris94} P.\ Avouris, I.\ Lyo, Science 264, 942 (1994). 
\bibitem{sanchez95} O.\ Sanchez, J.M.\ Garcia, P.\ Segovia, J.\ Alvarez, A.L.\ Vazquez de Parga, J.E.\ Ortega, M.\ Prietsch, R.\ Miranda, Phys.\ Rev.\ B52, 7894 (1995). 
\bibitem{garcia95_APA} J.M.\ Garcia, O.\ Sanchez, P.\ Segovia, J.E.\ Ortega, J.\ Alvarez, A.L.\ Vazquez de Parga, R.\ Miranda, Appl.\ Phys.\ A 61, 609 (1995). \bibitem{ortega00} J.E.\ Ortega, S.\ Speller, A.R.\ Bachmann, A.\ Mascaraque, E.G.\ Michel, A.\ N\"armann, A.\ Mugarza, A.\ Rubio, F.J.\ Himpsel, Phys.\ Rev.\ Lett.\ 84, 6110 (2000).
\bibitem{li97} J.\ Li, W.D.\ Schneider, R.\ Berndt, Phys.\ Rev.\ B56, 7656 (1997). 
\bibitem{li98} J.\ Li, W.D.\ Schneider, R.\ Berndt, S.\ Crampin, Phys.\ Rev.\ Lett.\ 80, 3332 (1998). 
\bibitem{jeandupeux99} O.\ Jeandupeux, L.\ B\"urgi, A.\ Hirstein, H.\ Brune, K.\ Kern, Phys.\ Rev.\ B59, 15926 (1999). 
\bibitem{buergi99} L.\ B\"urgi, O.\ Jeandupeux, H.\ Brune, K.\ Kern, Phys.\ Rev.\ Lett.\ 82, 4516 (1999). 
\bibitem{sprunger97} P.T.\ Sprunger, L.\ Petersen, E.W.\ Plummer, E.\ L\ae gsgaard, F.\ Besenbacher, Science 275, 1764 (1997). 
\bibitem{buergi00} L.\ B\"urgi, L.\ Petersen, H.\ Brune, K.\ Kern, Surf.\ Sci.\ 447, L157 (2000). 
\bibitem{repp00} J.\ Repp, F.\ Moresco, G.\ Meyer, K.-H.\ Rieder, P.\ Hyld\-gaard, M.\ Persson, Phys.\ Rev.\ Lett.\ 85, 2981 (2000). 
\bibitem{knorr02} N.\ Knorr, H.\ Brune, M.\ Epple, A.\ Hirstein, M.A.\ Schneider, K.\ Kern, Phys.\ Rev.\ {\bf 65}, 115420 (2002). 
\bibitem{morgenstern04} K.\ Morgenstern, K.-F.\ Braun, K.H.\ Rieder, Phys.\ Rev.\ Lett.\ {\bf 93}, 056102 (2004). 
\bibitem{morgenstern01} K.\ Morgenstern, F.\ Besenbacher, Phys.\ Rev.\ Lett.\ 87, 149603 (2001). 
\bibitem{morgenstern02} K.\ Morgenstern, K.-F.\ Braun, K.-H.\ Rieder, Phys.\ Rev.\ Lett.\ 89, 226801 (2002). 
\bibitem{fiete03} G.\ A.\ Fiete, E.\ J.\ Heller, Rev.\ Mod.\ Phys.\ 75, 933 (2003). 
\bibitem{fiete01} G.\ A. Fiete, J.\ S.\ Hersch, E.\ J.\ Heller, H. \ C. Manoharan, C. \ P. \ Lutz and D. \ M. Eigler, Phys. \ Rev. \ Lett. 86, 2392 (2001). 
\bibitem{everson90} M.P.\ Everson, R.C.\ Jaklevic, W.\ Shen, J.\ Vac.\ Sci.\ Technol.\ A8, 3662 (1990). 
\bibitem{hansemann03} M.\ Hansmann, J.I.\ Pascual, G.\ Ceballos, H.-P.\ Rust, K.\ Horn, Phys.\ Rev.\ B 67, 121409(R) (2003). 
\bibitem{baumberger04} F. Baumberger, M. Hengsberger, M. Muntwiler, M. Shi, J. Krempasky, L. Patthey, J. Osterwalder and T. Greber, Phys. \ Rev. \ Lett. 92, 196805 (2004). 
%\bibitem{dekker01} C.\ Dekker, et al.\ Nature 412, 617 (2001). 
%\bibitem{STM} K.-F.\ Braun, Ph.D. thesis, FU Berlin, 2001. 
%\bibitem{smith85} N.V.\ Smith, Phys.\ Rev.\ B32, 3549 (1985) 
%\bibitem{barral00} M.A.\ Barral, A.M.\ Llois, Phys.\ Rev.\ B62, 12668 (2000) % 2 
%\bibitem{williams78} R.S.\ Williams, P.S.\ Wehner, S.D.\ Kevan, R.F.\ Davis, D.A.\ Shirley, Phys.\ Rev.\ Lett.\ 41, 323 (1978) 
%\bibitem{heimann79} P.\ Heimann, H.\ Miosga, H.\ Neddermeyer, Phys.\ Rev.\ Lett.\ 42, 801 (1979) 
%\bibitem{shapiro88} A.P.\ Shapiro, T.\ Miller, T.-C.\ Chiang, Phys.\ Rev.\ B38, 1779 (1988) 
%\bibitem{davis85} R.F.\ Davis, R.S.\ Williams, S.D.\ Kevan, P.S.\ Wehner, D.A.\ Shirley, Phys.\ Rev.\ B31, 1997 (1985) 
%\bibitem{baumberger00} F.\ Baumberger, T.\ Gerber, J.\ Osterwalder, Phys.\ Rev.\ B62, 15431 (2000) 
%\bibitem{garcia95} N.\ Garcia, P.A.\ Serena, Surf.\ Sci.\ 330, L665 (1995) % 
%\bibitem{avouris94_jvstb} Ph.\ Avouris, I.-W.\ Lyo, R.E.\ Walkup, Y.\ Hasegawa, J.\ Vac.\ Sci.\ Technol.\ B12, 1447 (1994) % 
%\bibitem{davis91} L.C.\ Davis, M.P.\ Everson, R.C.\ Jaklevic, W.\ Shen, Phys.\ Rev.\ B43, 3821 (1991)% 
%\bibitem{kliewer01} J.\ Kliewer, R.\ Berndt, E.V.\ Chulkov, V.M.\ Silkin, P.M.\ Echenique, S.\ Crampin, Science 288, 1399 (2000). 
%\bibitem{fauster00} Th.\ Fauster, Ch.\ Reu\ss , I.L.\ Shumay, M.\ Weinelt, F.\ Theilmann, A.\ Goldmann, Phys.\ Rev.\ B 61, 16168 (2000) 
%\bibitem{echenique01} P.M.\ Echenique, J.\ Osma, M.\ Machado, V.M.\ Silkin, E.V.\ Chulkov, J.M.\ Pitarke, Prog.\ Surf.\ Sci.\ 67 (2001) 271 
%\bibitem{echenique00} P.M.\ Echenique, J.M.\ Pitarke, E.V.\ Chulkov, A.\ Rubio, Chem.\ Phys.\ 251 (2000) 1 
\end{thebibliography}
\end{document}